\documentstyle[12pt]{article}
\begin{document}

\centerline{ KNOT THEORY AND THE PROBLEM OF
 CLASSIFICATION }
\centerline{  FOR
PLASTIC STATE }
\bigskip
\bigskip
\centerline{\large Trinh Van Khoa}
\bigskip
\centerline{\large Department of Physics and ,}
\centerline{\large Moscow State University, 119899 Moscow, Russia}

\bigskip
\large

\bigskip
SUMMARY. In this paper we constructed new model of plastic deformation.
          The knot theory was used to classify the plastic state.\\

1. A long time ago there apppeared a paper of V.F.Jone [1,2,3] about
the knot
invariant. In this paper the connection between knot theory and
statistical mechanics was shown. It gives us new means and efficiency to
study the mechanical system in which appears dynamic symmetry in
evolutionary equations. In the most simple,  determining the knot
invariant is equivalent to determining the distribution at the
critical point in two dimension model of the phase transition. In
general, determining the knot invariant is equivalent to determining
the path
integral of the Feynman diagrams. It  mean that every Feynman diagram
 responds to one knot invariant. Moreover, symmetrical characters
of  Wu-Kadanoff-Wegner transformation are related to the topological
symmetrical characters of the knot in the model IRF for critical
process [4].

The Artin braid group Bn is the generalisation of the permutation
 group and is the
ingradient of the physical and mathematical theory in low dimension
manifold.The linear representation of group Bn is constructed in the
Hecke algebra of the group Sn. In the two dimension model the
integrability is connected to the subgroup of Bn in  Yang-Baster
equations . The group Bn is used in many papers. Here
 the method of group representation was used to construct Faynman
integral for probable distribution in the elastic-plastic
 problem.

2. Nowadays there are experimental data, allowing us to formulate a new
approach to plastic problems. It is the synergetic approach regarding
a solid as open , strong unequilibrium thermodynamic system and plastic
state - as a dissipative process [5,6]. According to this
approach plastic deformation of the solid may go on only at the
condition of dissimilar strain state. Plastic deformation originated
in the zone of  concentration of strain, as local phase transition
and expatiated only in the field of strain concentration as a relaxation
process.
The main strain concentrations are connected with the boundary
of the parts of medium or of defects. In general, the curve of strain
concentration (curve of defect) is the trajectory of deformation (Iliusin
trajectory of deformation). According to [7,8] deformation process
will be determined if deformation tensor will be given in the form of
function of time
$$ e=e(x,t ) $$
The process $e(x,t)$ was considered as the trajectory in a space.
Of course, it is the family of Iliusin trajectories. It means that it
has itself space as the group and representation corresponding to
it.

According to the approach of synergetics [9] the probability
of transition satisfies Fokker-Planck equation:
$$
\begin{array}{l}
  \frac{\partial}{\partial t}f(e,t)=- \frac{\partial}{\partial e}
  [K(e)f]+
  \frac{Q}{2}\frac{\partial^2}{\partial e^2}f
\end{array}\eqno (1)
$$
where $f(e,t)$ -  the probability of transition ,  e - order parameter
(in our models of plasticity it is tensor of deformation)[10],
Q - coefficient of diffusion, K- drift coefficient. Stationary
solution of Fokker-Planck
equation is given in the form
$$
\begin{array}{l}
 f(e)=M\exp[-\Phi(e)]
\end{array}\eqno (2)
$$
where $\Phi$ - takes the part of generalized thermodynamical
potential, M - constant.

 Non-stationary solution of Fokker-Planck
equation is given in the form of Feynman integral [11](for one dimension
and Q - irrespective of e )
$$
\begin{array}{l}
 f(e)=\lim\limits_{\begin{array}{l}
                    N\to \infty \\
                    N\tau=t
                    \end{array} }
\int\ldots \int De

 e^{-\frac{G}{2}}f(e',t_o)   \\

\begin{array}{l}

 De=(2Q\tau\pi)^{-N/2}de_o\ldots de_{N-1} \qquad e_N=e, e_o=e'\\

 G=\sum\limits_{\nu}\frac{\tau
 [\frac{e_{\nu}-e_{\nu-1}}{\nu}-k(e_{\nu-1})]^2}{Q} \\

\exp(-\frac{G}{2})=1-\frac12\int\limits_{t_1}^{t_2}Gf(e,t)de-
\frac{1}{2!}[\int\limits_{t_1}^{t_2}Gf(e,t)de]^2+\ldots
\end{array}
\end{array}\eqno (3)
$$

3. The basic idea of this approach is to take the knot as the
Iliusin's trajectory. We only consider the two dimension model. It is two
dimension complex manifold.
Space configuration of the system is given in form
$$
\begin{array}{l}
M_n=\{(Z_1,\ldots ,Z_n): Z_i\ne Z_j , i\ne j\}
\end{array}
$$
Homotopy class will be determined by winding number n. It is defined
for a piecewise differentiable curve C which does not pass through the
string b located at $\xi$:

$$
\begin{array}{l}
  n=\frac{1}{2\pi i}\int\limits_C\frac{dz}{z-\xi}
\end{array}\eqno (4)
$$
More generally, if we have chosen an origin and the two axes, we can
define the winding angle $\theta$ for the curve C with respect to the
b:
$$
\begin{array}{l}
\theta=\theta_1-\theta_o+2\pi n
\end{array}\eqno (5)
$$
where $\theta_o$  and  $\theta_1 $  are the angles of the start-point
and end-point of the curve respectively. According to [7,8]
 for every family of Iliusin
trajectories there is some internal space . Generically, we take the
internal space to be any Lie group. Thus we introduce the notion of
generalized winding number, which takes values in the internal space:
$$
\begin{array}{l}
\theta=sign(C)\mid\theta_1-\theta_o\mid+2\pi w \\
w=nT_a\otimes T_b=\frac{1}{2\pi i}\int\limits_C\frac{dz}{z-\xi}
T_a\otimes T_b
\end{array}\eqno (6)
$$
$T_a, T_b$ - the representation of string a and b.
  Let us parametrize the trajectory by z(t), $t_o\le t \le t_1$. Since
 we consider the Feynman propagator of string a going from $z_o=z(t_o)$
 to $z_1=z(t_1)$ . Drawing a parallel with [12] the constrained Feynman
 propagator of homotopy class $l$ for string a can be expressed formally
 as:
 $$
\begin{array}{l}
 K_l(z_1,t_1,z_o,t_o)=\int D_l z(t)D_l\overline{z}(t)
 \exp(i\int\limits_{t_o}^{t_1}Gdt)\delta^2(2\pi lT_a\otimes T_b-\theta)
\end{array}\eqno (7)
$$
here G is given in the solution (3) of Fokker-Planck equation. Dirac's
function can be represented by following Fourier transform:
$$
\begin{array}{l}
 \delta^2(2\pi lT_a\otimes T_b-\theta)
 =\int\frac{dkd\overline{k}}{4\pi^2}\exp[-i(k\phi+\overline{k}\phi)]
 \\
 P\exp\{i[2\pi k(lT_a\otimes T_b -w)]+c.c \}
\end{array}\eqno (8)
$$
It is a functional integral description of the topological properties
of the configuration space. Due to the string b, the space is no
longer simply connected. Because there  is self-organization in process
of plastic deformation, a non-trivial topological "interaction" arises
between string a and string b.  Substituting (8) into (7), we obtain
 $$
\begin{array}{l}
 K_l(z_1,t_1,z_o,t_o)=
                        \int\frac{dkd\overline{k}}{4\pi^2}
                       \exp[-i(k\phi+\overline{k}\phi)]  \\
 \qquad \qquad\qquad\qquad
 .\overline{K_l}(z_1,t_1,z_o,t_o;k,\overline{k})\\

  \overline{K_l}(z_1,t_1,z_o,t_o;k,\overline{k}) =
  \int D_l z(t)D_l\overline{z}(t)
 \exp(i\int\limits_{t_o}^{t_1}Gdt)\\
\qquad .\exp\{i\int\limits_{t_o}^{t_1}[k(\frac{iz^.}{z-\xi}+1\pi l)T_a\otimes
 T_b+c.c]dt \}

\end{array}\eqno (9)
$$
For  homotopy class $(l_1,\ldots,l_{i-1},l_{i+1},\ldots,l_n) $
denoting the difference in the initial angle and the final angle of
trajectory i with respect to trajectory j as $\phi_{ij}$:
 $$
\begin{array}{l}
\phi_{ij}=sign(C_i)\mid\theta_{ij1}-\theta_{ijo}\mid
\end{array}\eqno (10)
$$
we obtain the Feynman propagator for trajectory i carrying
represen-  tation $T_i$:
 $$
\begin{array}{l}
  K_{li}(z_{1i},t_1,z_{oi},t_o)=
                        \int\frac{dkd\overline{k}}{4\pi^2}
                       \exp[-i\sum\limits_{i,j=1}^{n}(k\phi_{ij}+
                       \overline{k}\phi_{ij})]
                       \\
 \qquad \qquad\qquad\qquad
 .\overline{K}_{li}(z_{li1},t_1,z_{lio},t_o;k,\overline{k})  \\

  \overline{K}_{li}(z_{i1},t_1,z_{io},t_o;k,\overline{k}) =
  \int D_l z(t)D_l\overline{z}(t)
 \exp(i\int\limits_{t_o}^{t_1}Gdt)\\
\qquad .\exp\{i\int\limits_{t_o}^{t_1}[k\sum_{j=1,j\ne}^{n}
(\frac{iz_i^.}{z_i-z_j}+1\pi l_j)T_i\otimes
 T_j+c.c]dt \}
\end{array}\eqno (11)
$$

4.  For every group Bn and it's representation, there is a partition
function . Topological structure of this group decides all characters
of the system. Then we can construct new model of plastic deformation
according to following scheme [10] :

First of all, Iliusin isotropy hypothesis will be understood: at the
beginning in the solid there are elastic strings or elastic region in
form of string

For every string there is a group and its representation

For every type of interaction of strings  there is one statistical sum or
its Feynman propagator, i.e there is the classification for plastic
state from knot invariant.

The probability of transfer to plastic state will be determined by
 Feynman integral

\end{document}